# Modified conformation and physical properties in conducting polymers due to varying conjugation and solvent interactions

Paramita Kar Choudhury,[*] Debjani Bagchi, C. S. Suchand Sangeeth, and Reghu Menon

Department of Physics, Indian Institute of Science, Bangalore − 560 012, India

Small angle X-ray scattering (SAXS) studies in poly[2-methoxy-5-(2'–ethyl-hexyloxy)-1,4-phenylene vinylene] (MEH-PPV) with varying conjugation, and polyethylene dioxythiophene complexed with polystyrene sulfonate (PEDOT-PSS) in different solvents have shown the important role of π-electron conjugation and solvent-chain interactions in controlling the chain conformation and assembly. In MEH-PPV, by increasing the extent of conjugation from 30 to 100 %, the persistence length ($l_p$) increases from 20 to 66 Å. Moreover, a pronounced second peak in the pair distribution function has been observed in fully conjugated chain, at larger length scales. This feature indicates that the chain segments tend to self-assemble as the conjugation along the chain increases. In case of PEDOT-PSS, the chains undergo solvent induced expansion and enhanced chain organization. The clusters formed by chains are better correlated in dimethyl sulfoxide (DMSO) solution than water, as observed in the scattered intensity profiles. The values of radius of gyration and the exponent (water: 2.6, DMSO: 2.31) of

---

[*] Electronic address: s_paramita@physics.iisc.ernet.in

Fax: +91-80-2360 2602

Tel: +91-80-2293 2859

power-law decay, obtained from the unified scattering function (Beaucage) analysis, give evidence for chain expansion from compact (in water) to extended coil in DMSO solutions, which is consistent with the Kratky plot analysis. The mechanism of this transition and the increase in dc conductivity of PEDOT-PSS in DMSO solution are discussed. The onset frequency for the increase in ac conduction as well as its temperature dependence probes the extent of connectivity in PEDOT-PSS system. The enhanced charge transport in PEDOT-PSS in DMSO is attributed to the extended chain conformation as observed in SAXS results.

## Introduction

It is well known that in conjugated polymers slight variations in structure and conformation can tune the physical properties significantly. Earlier studies have shown that intrinsic parameters like the extent of conjugation in polymer chains determine the π-electrons delocalization that in turn affects the rigidity and organization of the chains.[1] The consequence of these on the optical properties, in this regard, is of particular interest. The photophysical properties in semiconducting polymer like MEH-PPV[2] have shown that the nanoscale self-organization of chains affects the photoluminescence (PL), electro-luminescence and fluorescence properties.[3] For example, PL in PPV derivatives shows prominent red-shift due to the enhanced energy transfer processes as the π-electron delocalization increases.[4] This indicates that as the conjugation length increases, the π-π interactions can give rise to nanoscale aggregates; and in turn the PL spectra undergoes a red shift. Furthermore, the torsional angle between the benzene rings in polymer backbone tunes the conjugation,[1] and also the organization of chain segments via the interchain interactions, modify the fluorescence spectrum and PL quantum yield. In fluorescence studies, the quantum yield is observed to decrease when the interchain energy transfer increases. The earlier dynamic light scattering (DLS) study in PPV derivatives has shown that the variation in side group can influence the conjugation length, aggregation of chains and PL spectra.[5] Although these features in the optical phenomena have been observed, however there is a lack of structural studies to infer how the conformation and organization of chains could result in these properties. Hence more detailed studies regarding how conjugation and stiffness of chains affect the nanoscale organization of chain segments are required.

Moreover, the conformation and assembly of polymeric semiconductors can be modified by controlling the processing parameters like solvent, etc. An understanding of the role of chain conformation in solution state is important in device fabrication since the memory of the structural information in solution state is retained in solid films which can tune the physical properties of the devices.[6] Hence the correlation among the structural and physical properties of conducting and semiconducting polymers has gained lot of attention in recent years.[7] The SANS experiment in MEH-PPV has shown the role of solvents in nanoscale aggregation and also the formation of macroscopic nematic phase.[8] However, the primary conformation of chains in solution is yet to be investigated in detail to find out how the chain organization can be controlled by varying solvents and how it evolves in gels and films, especially for applications in organic electronics.[7] For example, PEDOT-PSS is a versatile system as it is soluble in water and several organic solvents.[9] PEDOT-PSS is considered as a promising material for replacing inorganic conductors in plastic electronics applications such as OLEDs, flexible photovoltaic devices, memories, and sensors, due to its ease of processing, high stability and transparency.[10,11] Usually, the appropriate combination of morphology and conductivity is required for various applications of PEDOT–PSS. The processing of PEDOT–PSS depends on a variety of factors including the amount of PEDOT, the fraction of solid content, the particle size and the viscosity which can lead to a wide range of morphological and electronic characteristics in dried films.[12] PEDOT–PSS dispersed in water can easily be spin-coated resulting in transparent films with high conductivity (typically 0.01 – 1 S/cm). The conductivity of films from aqueous solution can be enhanced by two orders of magnitude ~ 100 S/cm by mixing with glycerol, DMSO, etc.[13,14] This phenomenon was initially attributed to the screening effects of polar solvents on the Coulomb interaction between positively charged PEDOT grains and negatively charged PSS dopant, and also due to solvent-induced conformational changes in PEDOT-PSS chain, which controls the interchain interaction.[13] However, recent investigations indicate that this conductivity enhancement is mainly due to morphological changes in the film and the associated phase-segregation processes.[15-17]

In this work, the modification of conformation and organization of conducting polymer chains by varying conjugation and solvent are investigated. The role of the intrinsic parameter like conjugation and extrinsic parameter like solvent on physical properties of the polymer systems has been observed earlier. However, the underlying structural modifications subjected to such varying conditions with precisely quantified parameters have been systematically addressed in the present study. Towards this, small angle X-ray scattering (SAXS) studies in polymeric systems have shown that the conformation at various length scales can be probed to find out the correlations among the assembly of chains and nanomorphology.[18,19] SAXS studies in conducting polymer solutions are rather few, and although the earlier studies have given hints that the conformational modifications could play a role in the optical properties of MEH-PPV,[5] the direct structural investigations are lacking. While MEH-PPV has interesting optical properties, PEDOT-PSS has striking electrical properties that can be tuned by varying the structural parameters. The preliminary work in PEDOT-PSS has shown how the solvent can modify the chain conformation, however the structural aspects underlying this intriguing solvent induced effect are yet to be explored.[20] Although many works on electronic properties of conducting polymers have been reported, there are only a few that show how the variations in structural parameters in solution state are reflected in the electrical transport of solid films. This clearly shows the significance of the correlation among conformational and physical properties, which requires more detailed SAXS investigations that corroborates with the electrical conductivity measurements.

In the present work, SAXS measurements on the conformational modification of MEH-PPV due to the variation in conjugation as an intrinsic parameter is performed and the results are compared to the already known optical properties. The results show that by increasing the extent of conjugation from 30 to 100 %, the persistence length ($l_p$) increases from 20 to 66 Å. Moreover, a pronounced second peak in the pair distribution function observed in fully conjugated chain at larger length scales indicates that the chain segments tend to self-assemble as the conjugation along the chain increases. Furthermore, the SAXS studies in PEDOT-PSS solution have shown how extrinsic parameters like solvent can control the chain conformation that influences the electronic properties of the films, as revealed by dc as well as ac conductivity measurements. The results show that the chains get extended with the addition of DMSO, as compared to the aqueous solution of PEDOT-PSS. These results inferred from the intensity profile, Kratky plot and Beaucage analysis[21] show that the chains are flexible and the solvents can significantly change the nature of organization of the chains. A comparative study of dc and ac transport phenomenon in PEDOT–PSS in water and DMSO at low temperatures is carried out to investigate the network connectivity, correlated to the SAXS measurements.

**Experimental**

SAXS measurements were carried out using Bruker Nanostar equipped with a rotating anode source and three-pinhole collimation. A position sensitive 2D detector with 100 µm resolution was used to record the scattered intensity. The scattered intensity $I(q)$ is plotted as a function of the momentum transfer vector $q = 4\pi \sin\theta / \lambda$, where $\lambda$ is the wavelength of the X-rays (Cu-kα radiation, 1.54 Å), and $\theta$ is half the scattering angle. The $q$-range is 0.01 Å$^{-1}$ < $q$ < 0.3 Å$^{-1}$. The raw data was normalized for transmission coefficient, capillary width and exposure time; and also the incoherent scatterings due to solvent were subtracted in the data analysis.

The extent of conjugation (percentage of continuous alternating single and double bonds in a chain) in MEH-PPV samples are controlled by selective thermal elimination, as inferred from the absorption and photoluminescence peaks.[4] The sample with fully (100 %) conjugated chain i.e. without any break or discontinuity in conjugation, is named as M-PPV100 and the sample with much less (~ 30 %) conjugated chain is M-PPV30; with the corresponding absorption peaks at 508 and 425 nm, respectively.[4] Dilute solutions (1 % by weight) of these samples were prepared in xylene.

The second system PEDOT-PSS (short PEDOT chains complexed with PSS chain[11]) is an aqueous solution / dispersion of 1.1 wt.% of the polymer complex, was filtered with a 0.2 mm PTFE membrane filter, and then mixed thoroughly with DMSO in the ratio of 1 (pristine) : 3 (solvents) by volume. The sample solutions were filled in vacuum sealed quartz capillaries for the experiments. The chemical structures of MEH-PPV and PEDOT-PSS are shown in Fig. 1.

For the electrical measurements, PEDOT-PSS solution was mixed with 25% (by volume fraction) of DMSO. After stirring, the films were solution cast on glass substrate which was cleaned with triple distilled water, acetone, and isopropyl alcohol in ultrasonic bath for 20 min at each stage that is kept at 70 $^{o}$C for drying, and then annealed for 24 h to obtain free-standing films. Apart from this, the pristine aqueous PEDOT-PSS solution was drop cast to obtain the films that has lower conductivity. The transport measurements were performed using a standard four-probe dc technique, and conductive carbon paint was used to make contacts on the sample. The low temperature conductivity measurements are performed in a Janis continuous flow cryostat system. For dc conductivity measurements, current in the range 0.1–1 µA is applied (using Keithley 224 current source), and the voltage measured through Keithley 6514 electrometer. The ac conductivity measurements 40 Hz – 5 MHz, down to 5 K, were carried out using Agilent 4285A precision LCR meter and lock-in amplifier (SR-830), in Janis continuous flow cryostat.

**Results and discussion**

Figure 2 shows the scattering profiles of M-PPV100 and M-PPV30 solutions in xylene, and the inset shows the corresponding log I vs. log q plots. The intensity profiles follow a Debye function that decreases as $q^{-2}$ at small $q$ values, reflecting the coiled nature; and then follows a $q^{-1}$ behavior at larger $q$ as in rigid rods.[19] In both samples for $q$ > 0.1 Å$^{-1}$ the slopes of $I(q)$ are –1. In case of M-PPV100 the slope of $I(q)$ is –2 for $q$ < 0.1 Å$^{-1}$, and the shift in slope from -1 to -2 occurs at $q$ ~ 0.03 Å$^{-1}$. However, in M-PPV30 the change in slope from –1 to –1.3 is hardly discernable at $q$ ~ 0.1 Å$^{-1}$. The cut-off for the change in slope shifts to lower $q$ values as the extent of conjugation increases. The solutions are rather dilute (1 % by wt.) so that the scattering is mainly due to the form factor of well-separated chains. It is known that the slope -1 corresponds to the rigid-rod structure of chain segments.[19] This is in fact in accordance with the conjugation that can make the segments more rigid. At larger length scale ($q \leq 0.03$ Å$^{-1}$) the slope of $I(q)$ approaching -2 indicates presence of flexible coil like features in both systems. This is expected since semi-rigid polymer MEH-PPV is known to be constituted by hairy-rod segments where short coiled-segments are adhered to rigid backbone to enhance the solubility. However, in the sample M-PPV100, the slope change from -1 to -2 is rather abrupt as $q$ decreases; this is attributed to the formation of aggregates of rod-like segments that give rise to long-range organization, as also indicated from the optical studies.[4] This indicates that the extent of conjugation enhances the assembly of chain segments, and this is perceptible in other physical properties. These features are further analyzed to obtain more detailed structural parameters.

The fits in Fig. 2 follow the modified *worm-like chain* (WLC) model.[22-24] It deals with the local stiffness of chain segments at lower $q$ [parameterized by Kuhn length ($b$)], and also the Gaussian coiled structure at higher $q$ values; hence the model can be used to fit the data in the entire $q$-range, that also takes into account the excluded volume effect.[25] The solid lines are fits to Eq. 1 for this particular model[22] and the fit parameters are shown in the table in the inset of Fig. 2.

$$S_{exv}(q) = w(qR_g)S_D(q,L,b) + [1 - w(qR_g)][C_1(qR_g)^{-1/\nu} + C_2(qR_g)^{-2/\nu} + C_3(qR_g)^{-3/\nu}] \quad (1)$$

where $S_D = S_{Debye}(q,L,b) = 2[exp(-u) + u - 1]/u^2$ and $u = q^2R_g^2$ \quad (2)

The function $w(qR_g)$ is a crossover function:

$$w(x) = [1+ \tanh((x - C_4)/C_5)] \text{ with } x = qR_g \qquad (3)$$

The radius of gyration, $R_g$ is calculated by:

$$\langle R_g^2 \rangle = (Lb/6)[1 - (3/2n_b) + 3/2n_b^2 - 3/4n_b^3(1- \exp(-2n_b))] \qquad (4)$$

where $n_b$ ($=L/b$) is the number of Kuhn segments; $L$ is the chain contour length. The degree of rigidity as a measure of the curvature of chain can be estimated from the persistence length ($l_p$). The Kuhn length 'b' is related to $l_p$ by $2l_p=b$. The value of $L$ is calculated from $L = l_0 (M/M_0)$ where $l_0 = 6.7$ Å is the length of a monomer,[26] and $M$ and $M_0$ are the molecular weights[4] of polymer $M = 250000$ and monomer $M_0 = 276$. Using this relation, $L = 6068.8$ Å. In this analysis, although different structural length scales (from Kuhn segment to radius of gyration) are represented by different decades of $q$ in the intensity profile, the modified WLC Model can fit the entire range of $I$ vs. $q$, for both samples as shown in Fig 2. The features in the profiles indicate that the chains are represented by a combination of both flexible and rigid segments, as in the modified WLC model. The values of $R_g$ and $l_p$ for M-PPV100 in the inset-table can be compared to those from DLS (520 and 60 Å) and observed to be quite consistent.[5] It is interesting to note that $l_p$ for M-PPV100 is three times larger than that of M-PPV30, in accordance with the extent of conjugation. This implies that the fully conjugated chains in M-PPV100 are more rigid compared to the less conjugated M-PPV30. The enhanced π-π interaction of the rigid segments in former can favor the self-organization of chains, which is rather weak in latter. In fact this feature in the solution state is retained in solid films that in turn play a major role in the device characteristics.[27] The above structural features obtained from the data analysis in the Fourier space is further confirmed by the pair distribution function (PDF) analysis (to be discussed later) in the real space.

Earlier works in conjugated polymers have shown that the optical properties in PL and absorption spectra can be tuned by the polarity of solvent.[1,6] The SAXS data for MEH-PPV in dilute solutions of xylene (aromatic, non-polar) and THF (aliphatic, polar), have shown that the tendency to form aggregates increases in former solvent than in case of the latter, since non-aromatic and polar solvents like THF prefer solvating the side groups of MEH-PPV, while aromatic and non-polar solvents like xylene solvate the polymer backbone that give rise to different conformations.[28] Moreover, the small-angle neutron scattering (SANS) study in different solutions of MEH-PPV has shown the aggregation of chain segments to form nanoscale disc-like domains.[8] Many interesting solvent dependent optical as well as electrical properties have been observed in conjugated polymers,[6] which have been understood in detail with the help of our SAXS measurements in PEDOT-PSS solutions, and also compared to the conductivity measurements as described below.

The scattered intensity [$I(q)$], for PEDOT-PSS in water and DMSO solutions are shown in Fig. 3. The scattered intensity is lower in DMSO compared to water due to the dilution effect (the density of chain aggregates are lower, and the scattering is enhanced when the value of $q$ corresponds to the size of the aggregates). In both cases, $I(q)$ shows a substantial increase for $q < 0.03$ Å$^{-1}$. This typically occurs in presence of chain aggregates, and scattering is enhanced when the value of $q$ corresponds to the aggregate size.

Prior to the onset of this rapid increase in scattering, a detailed investigation of the profile shows the presence of a broad peak-like feature (absent in water, Fig. 3) at $q \sim$ 0.0279 Å$^{-1}$ for PEDOT-PSS in DMSO, which suggests that there is some degree of self-organization among the chain aggregates. The chains in DMSO therefore adopt a conformation that minimizes the internal energy of the system, so in spite of the fact that the entropy is reduced by this quasi-ordering, the free energy of the system is still minimized. The details of the solvent-chain interactions are explained after analyzing the single-chain conformation.

The conformational variation of chains in solution can be analyzed from the Kratky plot [$q^2I(q)$ vs. $q$], as shown in the inset of Fig. 3. Typically, tight compact coils exhibit a peak in Kratky profile; whereas in Gaussian random coils, $q^2I(q)$ has a short plateau at intermediate $q$ for coiled parts of the chain [$I(q) \approx 1/q^2$], followed by a linear increase at high $q$ (due to the contributions from short stiff segments of the chain [$I(q) \approx 1/q$]).[29,30] In both cases, the Kratky profiles do not show the rod-like limit even at higher $q$ values, so the chains are quite flexible, with short persistence length ($l_p$). The Kratky plot for aqueous solution has a broad peak, having a short plateau with a rapid downturn, indicating the presence of compact coil structure, though not as compact as in a globule. However, the Kratky-profile of PEDOT-PSS chains in DMSO solution has a broad plateau with a weak downturn at higher $q$ values (inset in Fig. 3). The extent of the plateau indicates up to what length scale the local structural feature persists in the system. This is longer in DMSO ($0.057 < q < 0.14$ Å$^{-1}$), implying a greater degree of intra-chain structural order. This local feature induces quasi-ordering among nearby chain segments, as observed in the smeared out shoulder in the scattering profile (Fig. 3).

The above qualitative conformational features are further analyzed in detail to obtain more quantitative structural parameters by using Beaucage's unified scattering function.[21] This method is usually used to study complex morphologies like polymeric mass fractals (fractal dimension between 1 and 3; 1 refers to a rigid rod-like structure, whereas 3 refers to a compact structure; an ideal Gaussian random coil's dimension is 2). Usually polymeric systems have different structural length scales (from Kuhn segment to radius of gyration) represented by power-laws [$I(q) \approx q^{-d}$, where $d$ is the dimensionality] for different decades of $q$ in the intensity profile. The log-log plot of $I(q)$ vs. $q$ (Fig. 4) shows a linear regime (power law behavior, extending about a decade in $q$) for $q > 0.04$ Å$^{-1}$, which evolves to the exponential regime at lower $q$ values. Hence, we have used Beaucage's unified equation (Eq. 5) to fit the intensity profiles for PEDOT-PSS in water and DMSO.

$$I(q) = G \exp\left(\frac{-q^2 R_g^2}{3}\right) + B\left(\frac{1}{q^*}\right)^p \tag{5}$$

where, $B = (pG/R_g^p)\Gamma(p/2)$ for a polymer mass fractal, and $q^* = q/[erf(kqR_g)/\sqrt{6}]^3$, ($k$ is considered to be 1.06 for a fractal dimension between 1.5 to 3),[21] $p$ is the power law exponent for the fractal dimension of the chains, $G$ is the Guinier pre-factor and $R_g$ is the radius of gyration of the chains. A non-linear least-square fit of the data with the unified equation is carried out using standard Levenberg-Marquardt algorithm, with statistical errors for the data points. The data fit quite well to Eq. 5 (Fig. 4) except at very low $q$; and the deviations from the fit for $q < 0.022$ Å$^{-1}$ is due to the presence of aggregates. The results of the fit along with reduced $\chi^2$ are compiled in inset-table of Fig. 4. The power law exponent for $I(q)$ profiles for these two systems reveal that PEDOT-PSS has a compact structure in aqueous solution ($p = 2.6$); whereas in DMSO solution, it is more elongated ($p = 2.3$). This is also evident from the $R_g$ values of PEDOT-PSS chains in water and DMSO solutions (Fig. 4 inset-table).

In the above analysis, though the role of solvent-induced conformational modification of chains is obvious, several competing electrostatic interactions play a crucial role in the expansion of chains in DMSO. So it is quite imperative to determine the Bjerrum length ($l_B$) and the Debye screening length ($\kappa^{-1}$). The value of $l_B$ is a measure of the strength of electrostatic interactions in different solvents, and it can be obtained from $l_B = e^2/4\pi^2\epsilon\epsilon_o k_B T$ (where $e$ is the charge of an electron in Coulombs, $\epsilon_o$ is the permittivity of free space, $\epsilon$ is dielectric constant of the solvents, $k_B$ is Boltzmann constant, $T$ is temperature). The values of $\epsilon$ for water and (water + DMSO) are 77.7 and 66.7, respectively.[31,32] The calculated values for $l_B$ are 7.16 and 8.42 Å for water and DMSO solutions, respectively; so the electrostatic interactions are stronger in DMSO. The value of $\kappa^{-1}$ is calculated from $\kappa^2 = 4\pi\, l_B\, C_S$ (where $C_S$ is the concentration of the counterions in moles/liter). In this case the counterions dissociated from PSS are H$^+$, and the average values of $C_S$ are 0.03 and 0.0016 moles/litre for water and DMSO solutions, respectively, as estimated from the pH values of the solutions. From this, the values of $\kappa^{-1}$ are 0.61 and 2.45 Å for water and DMSO solutions, respectively; implying that electrostatic interactions are short-ranged, though the range is considerably longer in DMSO solution.

Furthermore, $\kappa^{-1}$ can give a rough estimate for persistence length ($l_p$), which is a sum of the bare persistence length ($l_o$) and the electrostatic persistence length ($l_e$). The bare persistence length is mainly due to the rigidity of the chemical bonds and is calculated to be 7.8 Å for both the cases. The electrostatic persistence length for this flexible polymer in different solvents can be calculated according to the theoretical approach of Manghi and Netz[33] by a comparison of the screening length ($\kappa^{-1}$) and the electrostatic blob size given by $\xi \sim l_o (l_B \, l_o / A^2)^{-1/3}$ where A is the average separation between the charges in a chain (4.45 Å for PEDOT-PSS in both the solvents). For $\kappa\xi > 1$, screening is very large, and the electrostatic persistence length loses significance, i.e. $l_e = 0$. For $\kappa\xi < 0.1$, i.e. in case of weak screening, $l_e = 1/\kappa^2\xi$ and for $0.1 < \kappa\xi < 1$, $l_e \cong 1/\kappa$.[34,35] In the case of aqueous solution, $\kappa\xi > 1$, implying that $l_e = 0$, and persistence length is merely $l_o = 7.8$ Å. In DMSO solution, $\kappa\xi < 1$, so $l_e \cong 1/\kappa \cong 2.45$ Å. Hence the effective persistence length of PEDOT-PSS in DMSO solution is about 10.25 Å ($l_o + l_e$). It is therefore not surprising that the persistence regime (rod-like behavior of the chains) is not observed in the scattering profiles of both cases. Since $l_p$ is larger in DMSO solution the chains are less flexible with respect to chains in water, which could affect the short-range conformational details.

To further confirm the above data analysis, for both MEH-PPV and PEDOT-PSS, the pair (distance) distribution function $p(r)$ [the probability distribution of the vectors between pairs of scatterers] is calculated by inverse Fourier transform of the scattered intensity I(q), by using the algorithm GNOM[36]:

$$P(r) = \frac{1}{2\pi^2} \int_{q_{min}}^{q_{max}} I(q) qr \sin(qr) dq \qquad (6)$$

The average size of the chain $<R_g>$ can be calculated from the second moment of $p(r)$:

$$<R_g>^2 = \frac{\int_0^{r_{max}} r^2 p(r)dr}{2\int_0^{r_{max}} p(r)dr} \tag{7}$$

The $p(r)$ vs. $r$ (distance between a pair of scatters) data show the structural features present in both MEH-PPV and PEDOT-PSS samples, as in Fig. 5. The arrows show the corresponding axes for each sample. A comparison of the real-space values of $R_g$ and $l_p$ from this analysis are shown in Table I, and the values are quite similar to those obtained from WLC (MEH-PPV) and Beaucage (PEDOT-PSS) fits. MEH-PPV and PEDOT-PSS are quite different types of conducting polymer systems. MEH-PPV has a regular conjugated backbone while in PEDOT-PSS the short conjugated PEDOT segments occupy only a small portion of the flexible PSS chain. However, PDF profiles for both shows very similar features with the peak positions for MEH-PPV being shifted to larger length scales showing the presence of long-range order.

The $p(r)$ vs. $r$ data for PEDOT-PSS in two different solvents: water and DMSO, has two peaks in the range 0 – 350 Å which indicates the presence of short-range order. The first peak corresponds to intra-chain correlations among the segments, with the corresponding maximum in $r$ giving the most probable distance for the scatterers along the chain contour length, which consists of several Kuhn segments. The second peak accounts for inter-chain correlations and the corresponding maximum in $r$ indicates the distance to which the inter-chain segments can be correlated to each other.[29] The nearly symmetrical bell-shaped intra-chain correlation peak (around $r_1 = 33$ Å) in water shows globular compact coil shape of the chain. On the other hand, the asymmetry and weaker decay of intra-chain correlation peak of $p(r)$ of PEDOT-PSS in DMSO solution above 40 Å suggest the presence of expanded chains in DMSO [$p(r)$ is maximum at $r_1 = r_{max}/2$ for a spherical particle and this ratio of $r_1/r_{max}$ decreases as the chain elongates].

In case of M-PPV100 and M-PPV30 samples, the first peak at 175 Å corresponds to intra-chain and the second one at 700 Å is for long-range inter-chain correlations. The values of $R_g$ and $l_p$ from PDF analysis are given in Table I. These values agree very well with the parameters obtained from the fit to the modified WLC model, as in inset-table of Fig. 2. It is interesting to note that the second peak is quite pronounced for M-PPV100, showing that the organization of chain segments (due to enhanced π-π interactions) is considerably more with respect to the less conjugated sample. Moreover, the shifting of the second peak towards larger $r$, as the conjugation increases, indicates that the size of ordered domains increases. This implies the formation of a nematic-like phase due to the long range assembly of segments which is found to be in agreement with the features observed in optical properties too. The absorption, fluorescence and PL spectra are observed to undergo a red-shift as the conjugation length increases.[4] The SAXS data show that the second peak in PDF, particularly in M-PPV100, is due to the correlation between the assembly of extended and highly rigid chain segments. In very dilute solutions of polymers with varying conjugation lengths, interchain singlet exciton are "funneled" to lower energy states by energy transfer; the extent of which increases as the probability of finding longer conjugated segments in the nearest neighbourhood increases.[37] This directional energy transfer is attributed to the highly ordered conformation in the chains which explains the red-shift in both absorption and emission spectra in highly conjugated polymers. This interchain energy transfer further increases in films resulting in more prominent shifts in the spectra.

The above analysis shows that the conformational modifications and organizational processes in conjugated polymers can be generalized to a broader perspective. The length scales might vary indicating short (PEDOT-PSS) or long- range (MEH-PPV) order while the basic mechanisms behind the structural changes remain the same which get reflected in the electronic, optical or mechanical properties. Although the conjugation and solvent-dependent conformation are known to play a major role in the optical properties of MEH-PPV, the underlying structural features such as assembly of chain segments are revealed from the SAXS studies. On the other hand, the solvent-assisted variation in conductivity in PEDOT-PSS can be understood from the organization of chain segments as obtained from the SAXS data.

The temperature dependence of normalized resistivity ($\rho/\rho_{300\,K}$) for PEDOT-PSS in water, DMSO and PEDOT doped with $PF_6^-$ is presented[38] in Fig. 6. The $PF_6^-$ doped PEDOT system shows metallic behaviour, whereas the other two systems are in the insulating regime due to the hopping transport. The temperature dependence of resistivity of PEDOT-PSS in water is strong especially at low temperatures and becomes rather weak upon the addition of DMSO. In case of PEDOT-PSS in water, the resistivity increases by more than five orders of magnitude at low temperatures, whereas in DMSO it increases only by two orders, as shown in Fig. 6. The compact structure of the polymer chains in water attain an expanded-coil configuration with addition of DMSO which facilitates stacking of the chains because of enhanced interchain interaction. The localized charge carriers in compact coil (water solution) get delocalized and the chains become more extended upon the addition of DMSO, as already shown by the SAXS data. These delocalized carriers improve the interchain transport; as a result the temperature dependence of conductivity becomes weaker.[35] In charged conducting polymers like PEDOT-PSS the mobile charge carriers like polarons and bipolarons mainly govern the conduction mechanism.[39] A polaron corresponds to a positive charge on a monomer while a bipolaron corresponds to two positive charges delocalized over a few monomer units. This clearly indicates that polarons control the charge transport for the coiled chains in water, whereas the delocalized bipolarons dominate the transport in case of extended chain conformation in DMSO solution.

Although the dc conductivity of PEDOT-PSS shows enhanced delocalization of carriers upon the addition of DMSO, it is good to reconfirm this finding from the frequency dependence of conductivity. The frequency dependence of normalized ac conductivity of PEDOT-PSS in water and DMSO, at various temperatures, is shown in Fig. 7. The conductivity remains nearly constant at all the temperatures, up to a characteristic frequency called the onset frequency $\omega_0$, and increases $\sim \omega^s$ (with s < 1) at higher frequencies as observed in many disordered systems.[40,41] In this system the onset of frequency is $\sigma(\omega_0) = 1.1\sigma_0$, where $\sigma_0$ is the dc conductivity. For any disordered system, a correlation length $\lambda$ can be defined and this length scale corresponds to the distance between connections (e.g., junctions, nodes, etc.) in the system.[42] At the onset frequency, carrier travels a distance $\lambda$. It has been shown that, for disordered systems, the onset frequency scales inversely to some power of this correlation length. A higher value of the onset frequency implies smaller correlation length and shorter connections within the system.[12,43] The PEDOT-PSS in DMSO has superior transport characteristics in comparison to PEDOT-PSS in water, as observed from the significantly higher value of the onset frequency at 300 K (0.9 MHz for aqueous and 3.2 MHz for DMSO samples). At low temperatures, the ac conductivity is strongly frequency dependent, and the data show onset frequencies shift to lower values as temperature decreases. In aqueous PEDOT-PSS solution, the strong frequency dependence of ac conductivity indicates the weakening of the thermally-activated transport across the coiled up insulating PSS lamellae intervening between the conducting PEDOT-rich grains;[16] and upon the addition of DMSO the chain organization gets extended, thereby facilitating more efficient carrier transport across the barriers.

## Conclusions

In summary, the SAXS studies in MEH-PPV show modifications in chain conformation and assembly as the extent of conjugation and π–electron delocalization in MEH-PPV samples vary. The persistence length increases by a factor of three as the conjugation increases from 30 to 100 %. The presence of a prominent second peak in M-PPV100, at larger length scales, shows the presence of long-range organization among the chains. This type of chain conformation and organization can also be controlled by the solvent-chain interaction, as observed in the SAXS studies of PEDOT-PSS solutions. The Kratky plot and Beaucage analysis show the transition from compact (in water) to extended coil in DMSO solution as quantified by the $R_g$ values, and also the flexibility of chains from the $l_p$ values. This transition from compact to extended coil due to DMSO is also being reflected in the enhanced dc conductivity with weaker temperature dependence. The onset frequency for the increase in ac conduction as well as its temperature dependence directly probes the extent of connectivity in PEDOT-PSS. The enhanced charge transport in PEDOT-PSS in DMSO is governed by the extended chain conformation, unlike the compact coil chain structure in water. These results probe the conformation, interchain interactions and aggregation in the solution phase that affects the physical properties in solid films. This kind of systematic approach to understand the fundamental aspects behind the observed physical properties of conjugated systems has not been reported so far, to the best of our knowledge.

## Acknowledgments

P.K.C thanks Prof. S. Ramakrishnan for the MEH-PPV samples. C.S.S.S thanks CSIR, New Delhi for financial assistance.

*Oxford Univ. Press.*

20    D. Bagchi, and R. Menon, *Chem. Phys. Lett.* 2006, **425**, 114.

21    G. Beaucage, *J. Appl. Cryst.* 1996, **29**, 134.

22    J. S. Pedersen, and P. Schurtenburger, *Macromolecules* 1996, **29**, 7602.

23    O. Kratky, and G. Porod, *Rec. Trav. Chim. Pays-Bas* 1949, **68**, 1105.

24    P. Sharp, and V. A. Bloomfield, *Biopolymers* 1968, **6**, 1201.

25    M. Doi, S. F. Edwards, *The Theory Of Polymer Dynamics, Clarendon Press, Oxford*.

26    S. H. Chen, A. C. Su, Y. F. Huang, C. H. Su, G. Y. Peng, and S. A. Chen, *Macromolecules* 2002, **35**, 4229.

27    X. Deng, L. Zheng, C. Yang, Y. Li, G. Yu, and Y. Cao, *J. Phys. Chem. B* 2004, **108**, 3451.

28    P. Kar Choudhury, D. Bagchi, and R. Menon, *J. Phys.: Condens. Matter* 2009, **21**, 195801.

29    O. Glatter, and O. Kratky, *Small-Angle X-Ray Scattering (Academic Press, London, 1982)*.

**FIGURES**

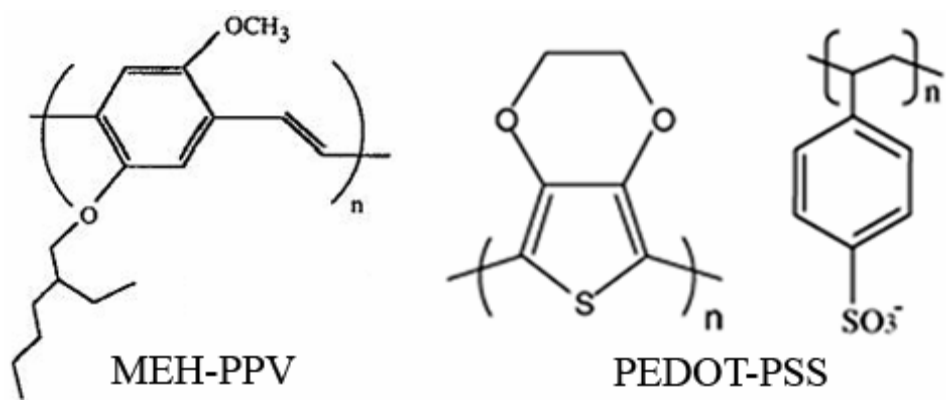

Fig. 1. Chemical structures of MEH-PPV and PEDOT-PSS.

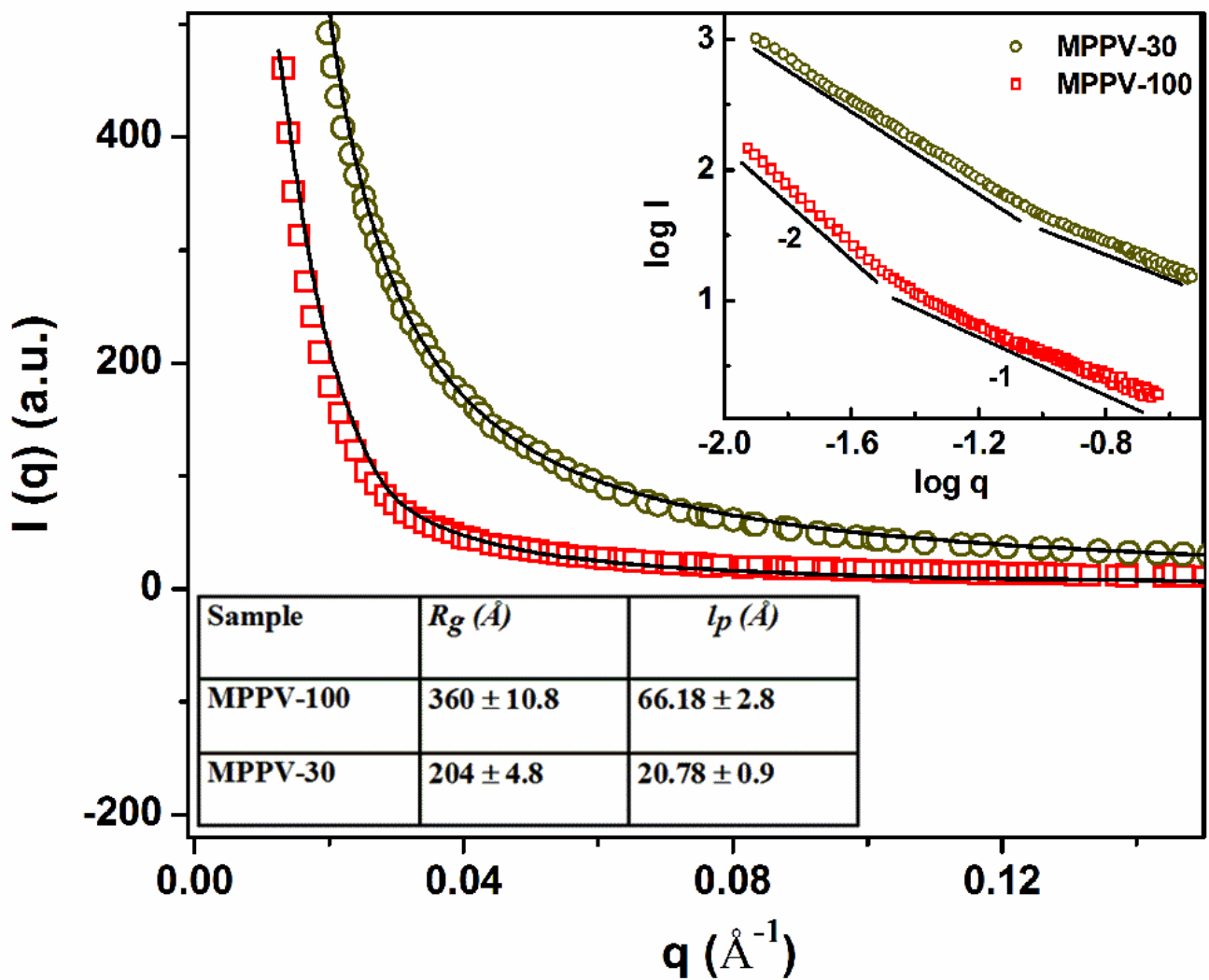

Fig. 2. Scattered intensity (I) vs. q for M-PPV100 and M-PPV30 solution in xylene. Solid lines are fits to worm-like chain (WLC) model (Eq. 1). Inset-table shows the fit parameters. Inset-figure shows the corresponding log I vs. log q plots of the scattering profile.

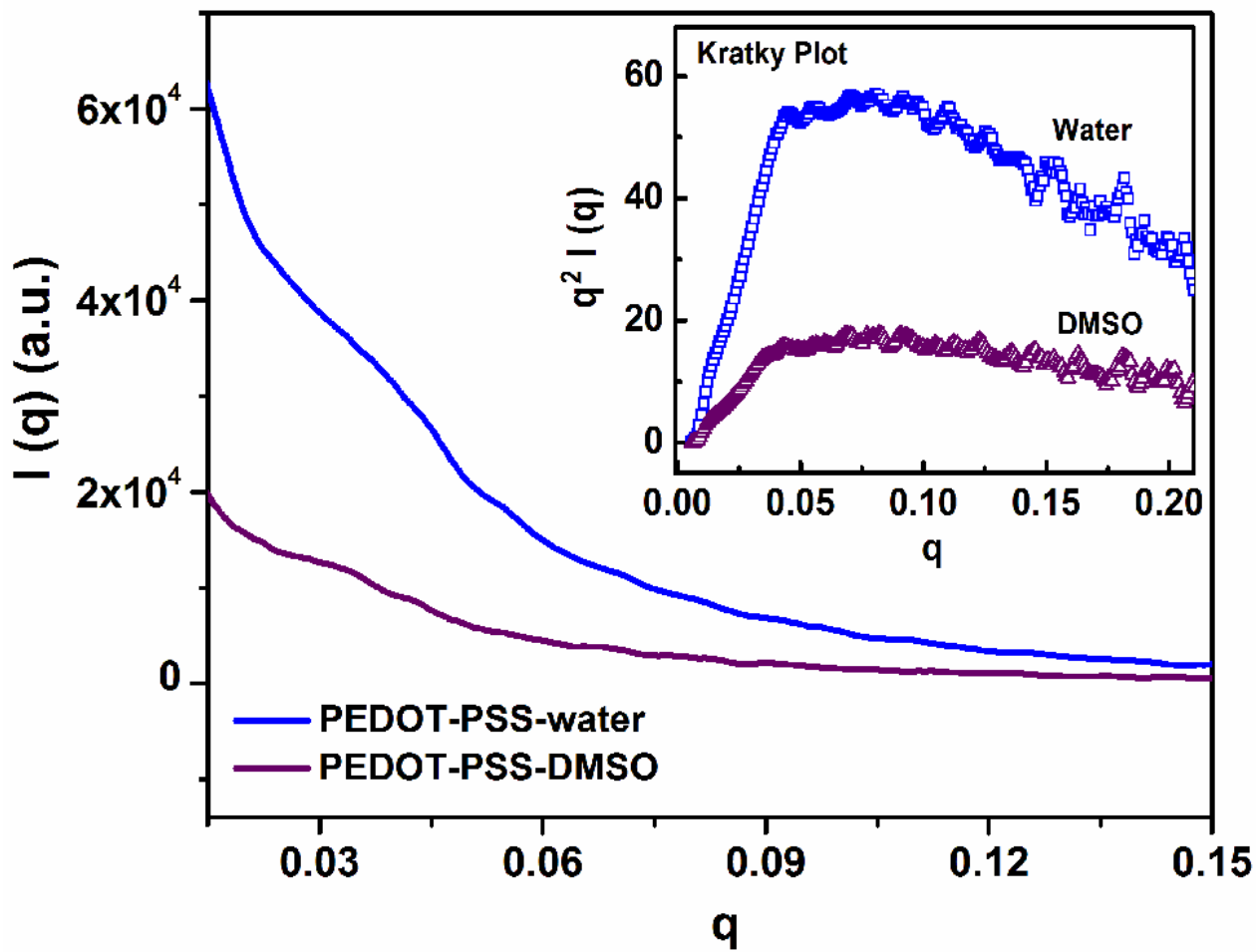

Fig. 3. Scattered intensity as a function of *q* for PEDOT-PSS solution in water and DMSO. Inset shows the corresponding Kratky plots [$q^2 I(q)$ vs. q]. A short plateau, with a rapid downturn from q = 0.1 for aqueous solution implies compact coils. The broad plateau, with a weak downturn at q = 0.14 for DMSO solution indicates extended coils, with more ordered features.

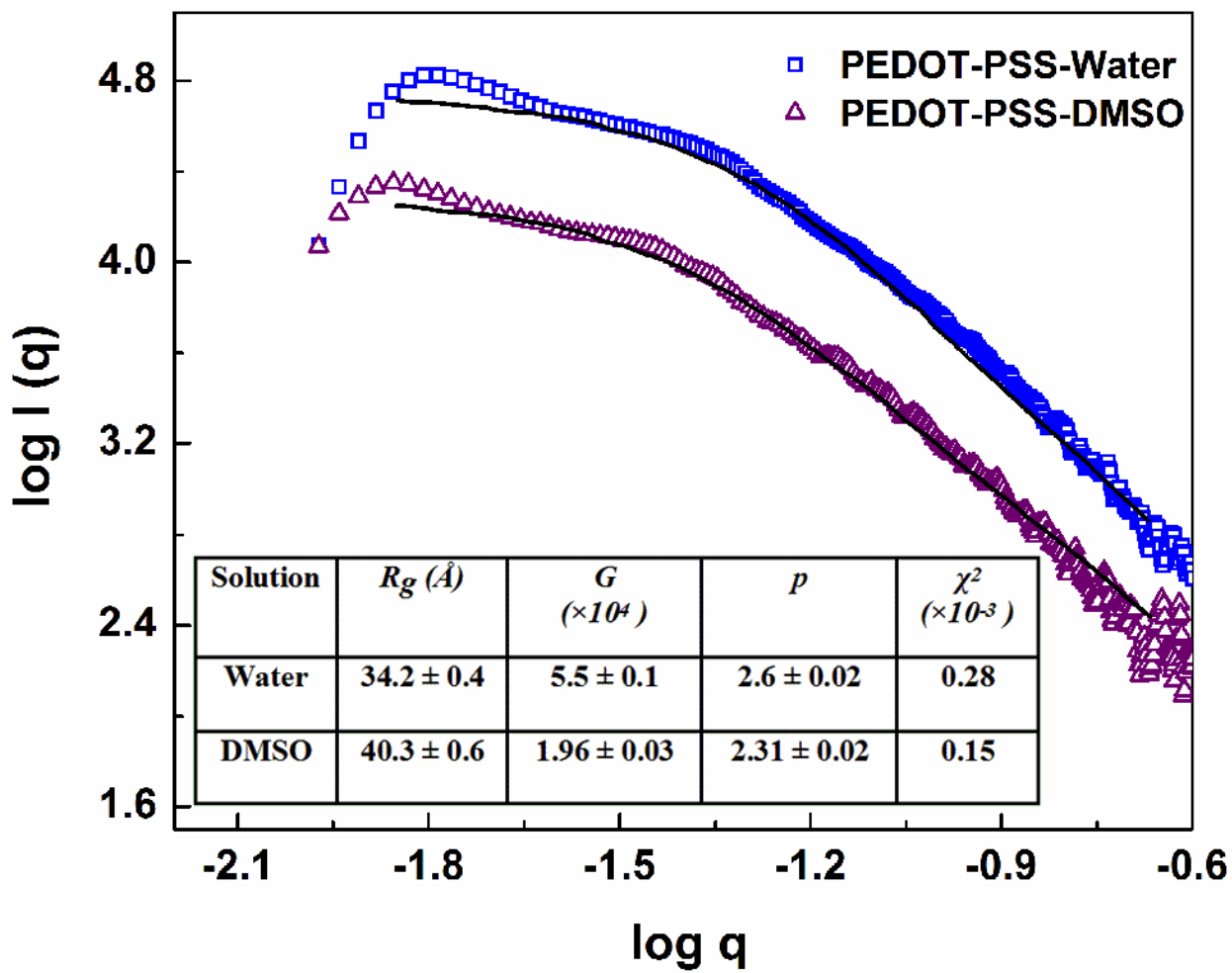

Fig. 4. Log-log plot of the scattered intensity as a function of q for PEDOT-PSS in water and DMSO solutions, and the linear region emphasizes power law behaviour. The solid lines are fits to Beaucage unified equation (Eq. 5). Inset-table shows the fit parameters.

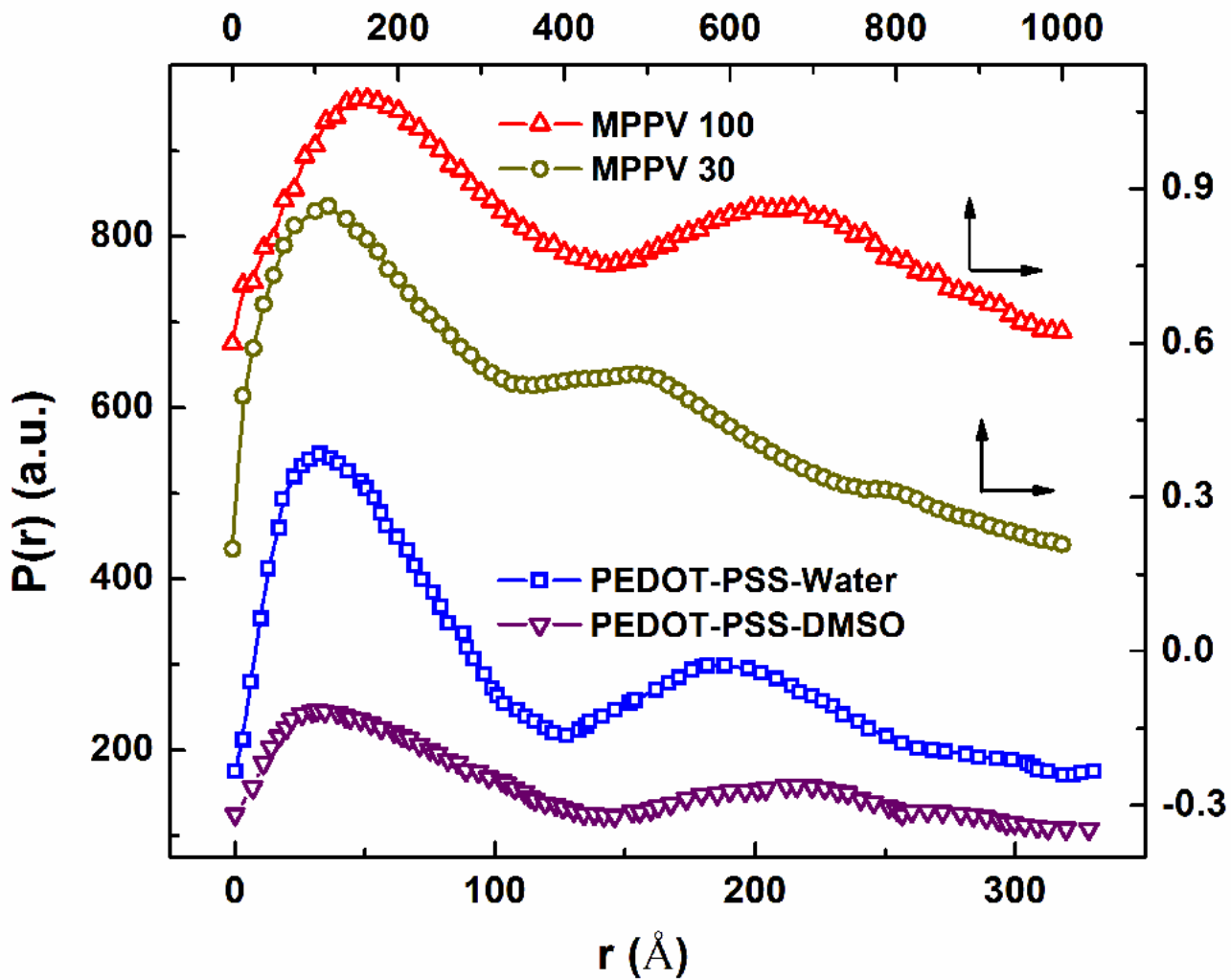

Fig. 5. Pair Distribution Function as a function of distance between scatterers (r). PDF for both MEH-PPV and PEDOT-PSS are compared, as indicated by the arrows to the axes. The baselines are shifted for clarity. The first and second peaks show the correlation among intra & inter-chain segments respectively.

31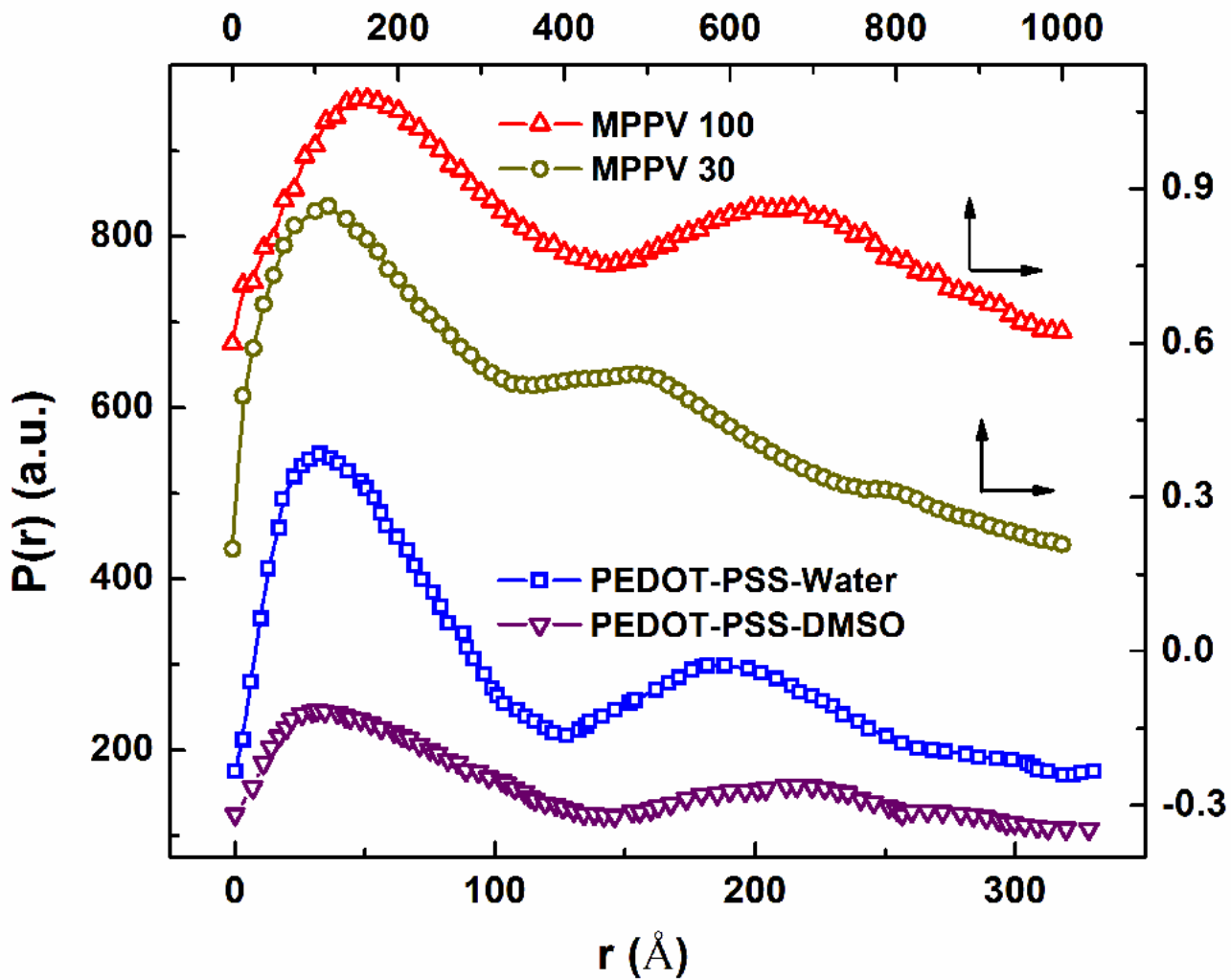

Fig. 5. Pair Distribution Function as a function of distance between scatterers (r). PDF for both MEH-PPV and PEDOT-PSS are compared, as indicated by the arrows to the axes. The baselines are shifted for clarity. The first and second peaks show the correlation among intra & inter-chain segments respectively.

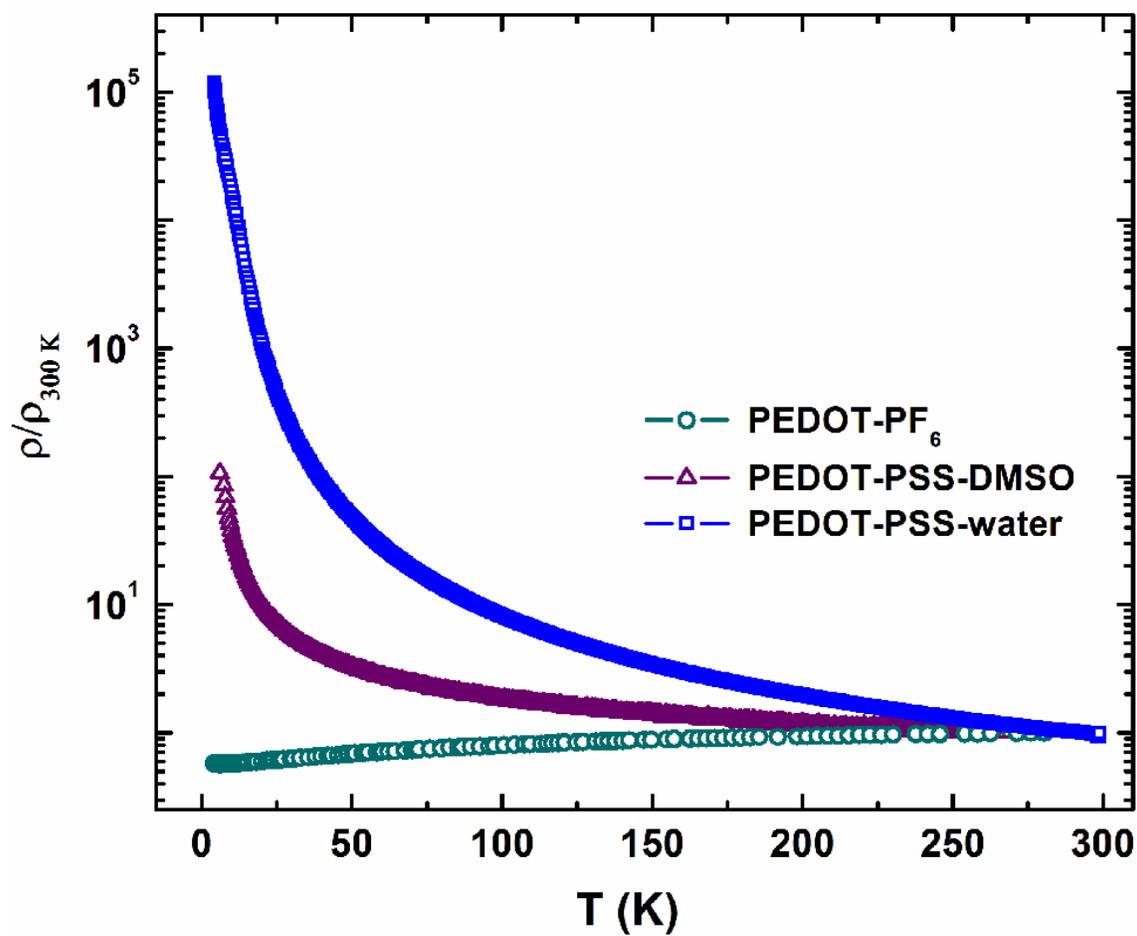

Fig.6. Temperature dependence of normalized resistivity of PEDOT-PSS in water and in DMSO solution, also compared with $PF_6^-$ doped PEDOT.

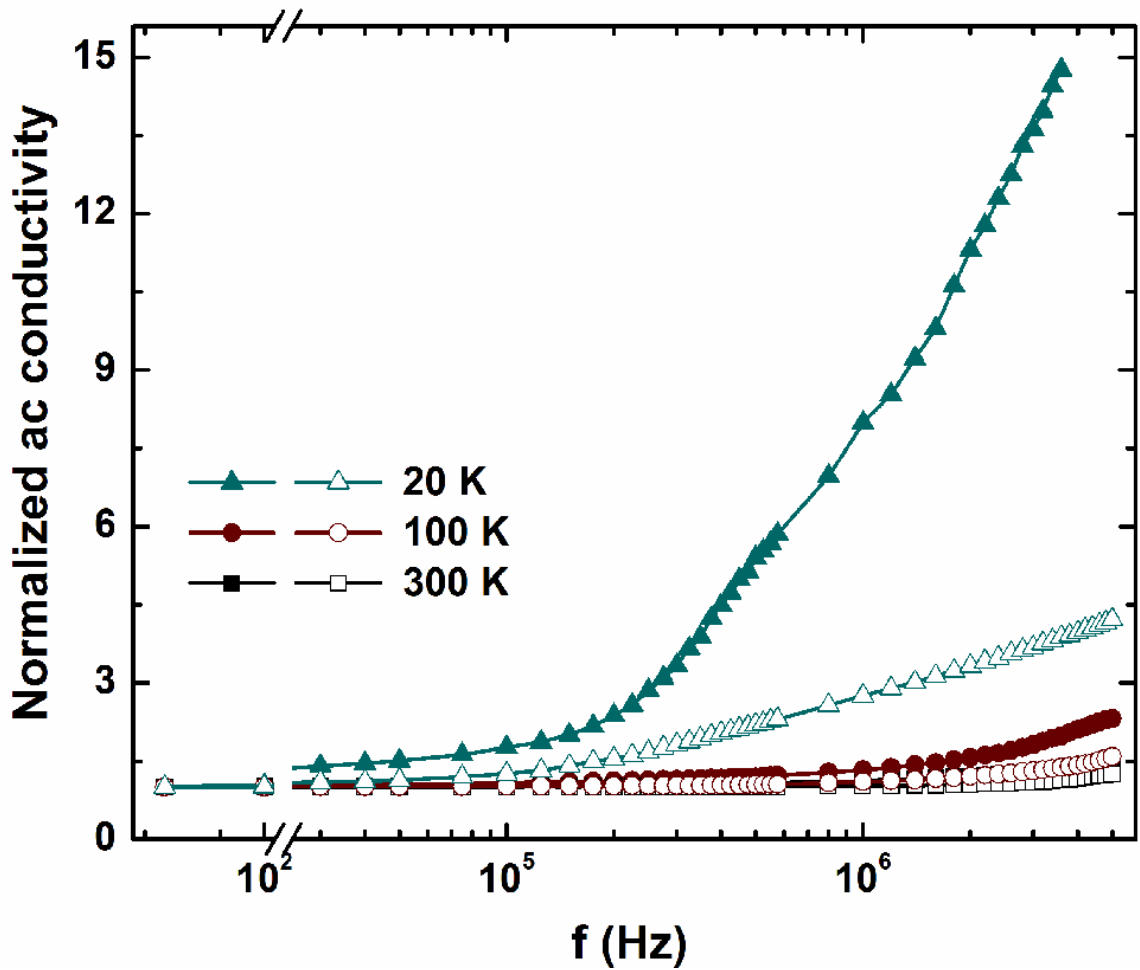

Fig. 7. Normalized ac conductivity of PEDOT-PSS in water (closed symbols) and DMSO (open symbols) at various temperatures.

Table I: Structural parameters [radius of gyration ($R_g$) and persistence length ($l_p$)] obtained from pair distribution function analysis (see Eq.s 6 and 7) for MEH-PPV and PEDOT-PSS samples.

| Sample | | $R_g$ (Å) | $l_p$ (Å) |
|---|---|---|---|
| MEH-PPV | 100 % | 336.25 ± 13.5 | 57.4 ± 3.3 |
| | 30 % | 209 ± 8.2 | 21.87 ± 0.67 |
| PEDOT-PSS | Water | 38.2 ± 0.22 | 7.8 |
| | DMSO | 44.4 ± 0.6 | 10.25 |